\documentclass{JHEP3} 

\usepackage{epsfig,multicol}
\usepackage{amsmath}

\newcommand\fverb{\setbox\fverbbox=\hbox\bgroup\verb}
\newcommand\fverbdo{\egroup\medskip\noindent%
            \fbox{\unhbox\fverbbox}\ }
\newcommand\fverbit{\egroup\item[\fbox{\unhbox\fverbbox}]}
\newbox\fverbbox

\title{Brane resolution in heterotic theory through BF term}

\author{J. E. G. Silva$^{a}$, F. A. Brito$^{b}$, and C. A. S. Almeida$^a$
\\
$^a$Departamento de F\'{i}sica - Universidade Federal do Cear\'{a} \\ C.P. 6030, 60455-760
Fortaleza-Cear\' {a}-Brazil
\\ \\
$^b$Departamento de F\'{i}sica, Universidade Federal de Campina Grande,
Caixa Postal 10071, 58109-970 Campina Grande, Para\'{i}ba, Brazil}

\abstract{We have studied the resolution of a naked singularity of a conifold in heterotic theory by a BF topological
defect living in a 5-brane. The singularity is removed due to Chern-Simons action that changes the Bianchi identity
for the $\mathcal{H}_{3}$ 3-form. Following the previous analysis of Cvetic, L\"{u} and Pope \cite{Cvetic:2000mh} where they have studied
the resolution through an instanton defect, we have taken a conifold over an \textit{Eguchi-Hanson} manifold and a harmonic self-dual
2-form related with $F_{2}$ to solve the differential equation for the warp factor. Since the $\mathcal{H}_{3}$ field
is related to torsion in the extra manifold, we can interpret this conifold as one with torsion. Using the so called
BF term we have found a solution with the same properties of the instanton such that the conifold is smoothed
out and has a torsion that diverges in IR regime and vanish in UV regime.}


\begin{document}


\section{Introduction}

Conifolds are manifolds of the form $C^{2n}=R^{+}\times X^{2n-1}$, where $X^{2n-1}$ is a manifold
topologically equivalent to a $S^{2n-1}$ \cite{Candelas:1989js}, with a metric of the form
$ds^{2}(C^{2n})=dr^{2}+r^{2}ds^{2}(X^{2n-1})$.
It is a kind of orbifold generated by the action of the group $Z_{n}$ over $R^{n}$ and therefore
has a naked singularity in $r=0$ due to a fixed point of the action of the group \cite{kms}. This point
is important because we can transform a kind of Calabi-Yau space into another one just collapsing a cycle and so
generating a conifold inside the extra dimensions, process called conifold transitions \cite{p}.

Another importance of conifolds lies in extensions of AdS-CFT correspondence \cite{Maldacena:2000yy}. In fact,
Klebanov and Strassler \cite{Klebanov:2000hb}
have studied a extension of this theory using the manifold $ \mathcal{M}^{10}=M^{4}\times C^{6}$ that asymptotically
is $AdS^{5}\times X^{5}$ instead of $AdS_{5}\times S^{5}$ ,
both in the type II-B
string theory. Supposing $X^{5}$ Ricci-flat, they analyzed the flux of renormalization group and concluded that the conformal field theory of dual fields
is preserved only if the naked singularity is smooth out. To do it, they have added fractional branes whose flux of
gauge fields due the Chern-Simons or anomaly term, removes the singularity in a way previously studied by
Candelas and de la Ossa\cite{Candelas:1989js} called deformation.  The deformation procedure geometrically deforms the quadric that defines
the conifold. Pando-Zayas and Tseytlin \cite{Pando Zayas:2000sq} have found another way of smooth the conifold introducing an asymmetry in $X^{5}$.
This does not alter the quadric but introduces a resolution parameter and so it is called resolution.

Cvetic, L\"{u} and Pope \cite{Cvetic:2000mh} have studied the resolution of a cone defined over an Eguchi-Hanson
space in a heterotic string theory. They have chosen a warped product of a 5-brane with an Eguchi-Hanson space and
taken an instanton defect living in Eguchi-Hanson space.  The choise of an Eguchi-Hanson space is due this
manifold be a  solution of Einstein equations of the K\"{a}hler Ricci-flat type  and also because it has a self-dual
Riemann curvature \cite{Eguchi:1978xp,Mahapatra:1999eb,Fu:2008ga,Singh:2007vw}. The instanton defect appears naturally in the anomaly action
in the supergravity regime altering the Bianchi identity for the 3-form
of Kalb-Ramond $\mathcal{H}_{3}$ field that can be understood as a magnet flux through the transverse space. It also
plays an
important role in AdS/QCD, where it is holographically related with barions \cite{Hong:2006ta,Hata:2007mb}. On the other hand, using the transgression of Bianchi identity,
it is found a differential equation for the warp factor depending on $F_{2}$. Also, a self-dual
harmonic 2-form $L_{2}$ such that $F_{2}=mL_{2}$ it is found. Then if $L_{2}$ is $L^{2}$ normalizable the singularity is
taken off \cite{Cvetic:2000mh}. The $L_{2}$ exists because the transverse space is Ricci-flat and it has a covariantly
constant spinor. Therefore, $L_{2}$ has a closed fundamental 2-form that also is a K\"{a}hler form \cite{Cvetic:2000db}.

Recently, Carlevaro and Isr\"{a}el \cite{Carlevaro:2009jx} have studied the resolution of a cone in a more wide scenario. They have
taken a non-warped product of a Minkowski space with a conformal 6-conifold over a $T^{1,1}$ arbitrary base space in a heterotic string theory
with a non-vanishing 3-form $\mathcal{H}_{3}$. Since we can define a torsion form from $\mathcal{H}_{3}$, this conifold is not a Ricci-flat space. So, in order to determine the geometry they solved the Strominger equation \cite{Strominger:1986uh}, necessary condition for supersymmetric invariance. This equation couple
the exterior derivative of a 2-fundamental form and a 3-holomorfic form to the dilaton. So, this manifold has a
varying dilaton
that makes the string coupling not a constant. Assuming the conifold is asymptotically Ricci-flat, a
gauge bundle is Abelian and taking the double scaling limit \cite{Carlevaro:2008qf} where the string coupling constant goes to zero and the
ratio $\frac{g_{s}\alpha'}{a}$ is fixed, the authors have found analytically and numerically a solution that is a torsional
analogue of solutions of Eguchi-Hanson space in six dimensions. Since the $\mathcal{H}_{3}$ diverges only at origin the torsion have
a IR singularity and vanish at infinity.

Torsional manifolds have being extensively studied because, on one hand, it can be viewed as spinning
branes that are solutions extending the black branes to include spin (Euclidean Kerr solution) and this generates torsion \cite{j}.
On another hand, torsional manifolds appear on study of compactifications with non-vanishing fluxes that explain the stabilization of moduli fields,
necessary condition for deduce the coupling constants of standard model \cite{Becker:2003yv}. Furthermore, torsion effects also
have importance in brane worlds scenarios where they alter the area of event horizon of four dimensional black holes \cite{HoffdaSilva:2009ht}.

Motivated by these two issues, namely, the resolution of conifolds and the inclusion of torsion in extra dimensions,
both due to instanton $F_{2}\wedge F_{2}$ term,
we have studied the resolution of conifold due a topological defect generated by a $B\wedge F$ term
instead of $F\wedge F$ one. On one hand this change is possible because both are 4-forms metric independent. Further,
the $B\wedge F$ belongs to the same Chern Class of $F\wedge F$ that appears in the anomaly term in type II string theories.
Since both $B_{2}$ and $F_{2}$ belong
to Neveu-Schwarz sector and heterotic supergravity does not have Ramond-Ramond fields, we propose to make the same
shift $\mathcal{F}_{2}\longrightarrow B_{2}+kF_{2}$ in Chern-Simons action and study the effects over the transgression
of Bianchi identity \cite{kms}. On another hand, the BF theory is
interesting by itself because it is an extension of 3D Chern-Simons theory to four dimensions and so it is a theory
for topological gravity \cite{Capovilla:1991kx}-\cite{Tahim:2005wg}. Further, since 3D Chern-Simons theory describes the topological models for fractional statistics
for planar systems, the BF theory can do the same for three space dimensions where on has interesting applications in condensed matter \cite{Almeida:2001nt,Medeiros:1999df}. Following the steps of Cvetic, L\"{u} and Pope \cite{Cvetic:2000mh}, we take a conformal conifold
over an Eguchi-Hanson space, with the same ansatz previously used for a $L_{2}$. Using a well known ansatz for $B_{2}$
form we obtain a differential equation for the warp factor. Now, we can tuning the strength of $B_{2}$ and $F_{2}$ to
find a solution that resolves the conifold and in addition has the same torsional properties of the resolved conifold
due the instanton term.

This work is organized as follows: in section II we obtain the BF and instanton term from the general Chern-Simons
action and we present the transgression of Bianchi identity.  We study the general characteristics of four
dimensional Euclidean
Ricci-flat spaces in section III whereas in section IV we review the resolution using the instanton term only over an Eguchi-Hanson conifold
as done in \cite{Cvetic:2000mh} discussing the behavior of warp factor and the torsion. In section V, we obtain a
resolved solution with the same features of instanton using the BF term instead of instanton one and finally, our conclusions are presented in the section VI.

\section{BF term in heterotic theory}

Consider the bosonic sector of heterotic theory in the supergravity regime whose action is given by \cite{kms,p,rs}:

\begin{equation}
 S_{10}^{het}=S_{NS}+S_{YM}+S_{CS},
\end{equation}
where $S_{NS}$ is the Neveu-Schwarz action for the curvature $R$, for the dilaton field ($\phi$) and for the NS-NS field strength
($\mathcal{H}_{3}$) and is expressed by
\begin{equation}
S_{NS}= \frac{1}{2\kappa^{2}}\int_{M_{10}}d^{10}x\sqrt{-g}e^{-2\phi}(R+4||\nabla \phi||^{2}-\frac{1}{2}||\mathcal{H}_{3}||^{2}).
\end{equation}
$S_{YM}$ is an action of Yang-Mills type of form

\begin{equation}
 S_{YM}= \frac{1}{2g^{2}}\int_{M_{10}}{tr (\star_{10}R_{2}\wedge R_{2} + \star_{10}F_{2}\wedge F_{2})},
\end{equation}
where the $R_{2}$ is the curvature 2-form and the $F_{2}$ is the field strength for the $A_{1}$ gauge field.
This term gives gravitational correction of higher order and introduces an instanton defect.

The two actions above are entire world-volume or metric dependent. However, for the cancelation
of anomalies we must add the anomaly or Chern-Simons action $S_{CS}$ given by

\begin{equation}
 S^{het}_{CS}=\int{B_{2}\wedge X_{8}(F_{2},R_{2})},
\end{equation}
where $X_{8}$ is the symmetric polynomial of eighth order both in gauge and Lorentz curvature. By the Green-Schwarz mechanism
there exists a 4-form $Y_{4}$ such that

\begin{equation}
 d\mathcal{H}_{3}=Y_{4}=tr(R_{2}\wedge R_{2}-F_{2}\wedge F_{2}).
\end{equation}
This is called a transgression of the Bianchi identity for the $\mathcal{H}_{3}$ field. In order to NS-NS field
strength
$\mathcal{H}_{3}$ satisfy this equation we must add to this field a topological term

\begin{equation}
 \mathcal{H}_{3}=dB_{2}+\frac{l^{2}_{s}}{4}tr\left(\omega_{1}\wedge d\omega_{1} + \frac{2}{3}\omega_{1}^{3}-(A_{1}\wedge dA_{1} + \frac{2}{3}A_{1}^{3})\right).
\end{equation}
This additional term is nothing but the Chern-Simons 3-form for the spin connection $\omega_{1}$ and for the
gauge field $A_{1}$.

Extremizing the action for $B_{2}$ field yields the equations:
\begin{eqnarray}
 d\star_{10} \mathcal{H}_{3} & = & 0\\ \label{X8}
 d\mathcal{H}_{3} & = & tr(R_{2}\wedge R_{2} -F_{2}\wedge F_{2}).
\end{eqnarray}
In \eqref{X8} we neglect higher order terms in $F_{2}$ and $R_{2}$.
These equations are essentials for the resolution of four-dimensional spaces.
Integrating the previous equation we obtain
\begin{eqnarray}
 \chi=\int_{M_{4}}{tr(R_{2}\wedge R_{2})}=\int_{M_{4}}{tr(F_{2}\wedge F_{2})},
\end{eqnarray}
where $\chi$ is the Euler number and the integral $\int_{M_{4}}{tr(F_{2}\wedge F_{2})}$ is the instanton charge.
Then we can interpret the resolution process as a result of the flux generated by a topological defect of instanton type
living in a 4-brane. This open the question: could be possible a resolution using another topological defect living in 4-branes?

In type II string theories the Chern-Simons action for the anomaly cancelation \cite{rs} is
\begin{equation}
 S^{II}_{CS}=\frac{T_{p}}{2}\int{tr\left(e^{\mathcal{F}_{2}}\wedge C\wedge \sqrt{\frac{\hat{A}(R_{T})}{\hat{A}(R_{N})}}\right)}
\end{equation}
where $\mathcal{F}_{2}=B_{2}+kF_{2}$, $\kappa=2\pi \alpha'=2\pi \sqrt{l_{s}}$ \cite{kms}.
Let us consider only the trivial cases where $\hat{A}(R_{T})=\hat{A}(R_{N})=1$. Expanding the exponential until second order we obtain $tr(B_{2}\wedge B_{2} + 2\kappa B_{2}\wedge F_{2} +
\kappa^{2}F_{2}\wedge F_{2})$. Note that this yields three different topological defects that, if we suppose them to live
in a 4-brane, we can obtain their topological charge by fluxes. We can interpret this expansion in terms of constant $\kappa$ and so the $B\wedge F$ term
is relevant in first order in $\kappa$, what means for a scale $l_{s}>l_{p}$, while the instanton term $F\wedge F$ is relevant
for $\kappa^{2}$, that means $l_{s}>>l_{p}$.

Inspired in the Chern-Simons term
in type II theories, let us make a replacement $F_{2}\rightarrow \mathcal{F}_{2}=B_{2}+kF_{2}$ in heterotic theory.
This change can be done because this field already belongs to the heterotic action. Further, although this change is based
in the anomaly term in type II theories, it does not depends on the R-R fields that do not exist in the heterotic theory.

By doing this change the Yang-Mills term in the action turns to be:
\begin{equation}
 S_{YM}=\int_{M_{10}}{tr (\star_{10}R_{2}\wedge R_{2} + \star_{10}B_{2}\wedge B_{2}+2k\star_{10}F_{2}\wedge B_{2}+k^{2}\star_{10}F_{2}\wedge F_{2} )}.
\end{equation}
The 4-form $Y_{4}$ should also be altered for
\begin{equation}
 Y_{4}=tr(R_{2}\wedge R_{2} -2kB_{2}\wedge F_{2}-k^{2}F_{2}\wedge F_{2}).
\end{equation}
Hence the NS-NS or Kalb-Ramond field strength must satisfies
\begin{equation}
d\mathcal{H}_{3} = tr(R_{2}\wedge R_{2} -2kB_{2}\wedge F_{2}-k^{2}F_{2}\wedge F_{2}).
\end{equation}
Then we must add another topological term for the $\mathcal{H}_{3}$ of the form
\begin{equation}
 \mathcal{H}_{3}=dB_{2}+\frac{l^{2}_{s}}{4}tr\left(\omega_{1}\wedge d\omega_{1} + \frac{2}{3}\omega_{1}^{3}-(A_{1}\wedge dA_{1} + \frac{2}{3}A_{1}^{3})-B_{2}\wedge A_{1}\right).
\end{equation}

This changes yields to new equations for $\mathcal{H}_{3}$, namely
\begin{eqnarray}
 d\star_{10} \mathcal{H}_{3} & = & X_{8}-\star_{10}B_{2}-\star_{10} F_{2} \\
 d\mathcal{H}_{3} & = & tr(R_{2}\wedge R_{2}-kdB_{2}\wedge A_{1} -2kB_{2}\wedge F_{2}-k^{2}F_{2}\wedge F_{2}).
\end{eqnarray}

Since we want to study the resolution through a BF term and compare with the instanton term approach let us consider only the
$B_{2}\wedge F_{2}$ and $F_{2}\wedge F_{2}$ terms in anomaly term. Further, we will disregard the
$\star_{10}R_{2}\wedge R_{2}$ and $\star_{10}B_{2}\wedge B_{2}$ terms. So we will study solutions of the system of equations
\begin{eqnarray}
 d\star_{10} \mathcal{H}_{3} & = & X_{8}-\star_{10}B_{2}-\star_{10} F_{2} \\
 d\mathcal{H}_{3} & = & tr(-2kB_{2}\wedge F_{2}-k^{2}F_{2}\wedge F_{2}).
\end{eqnarray}
The transgression of Bianchi identity represents the magnetic flux due a instanton and a
BF topological defect. The 2-form field strength usually take values on
$SU(32)$ or $E_{8}\times E_{8}$ gauge group but we will consider only an Abelian $U(1)$ case for simplicity.

\section{Resolution over four-dimensional transverse spaces}

As previously done by Cvetic, L\"{u} and Pope \cite{Cvetic:2000mh}, let us take the bulk as the warped product between
a D5 brane
and a Ricci-flat manifold:

\begin{equation}
 \mathcal{M}^{10}=R^{1,5}\times M^{4},
\end{equation}
where $M^{4}$ is a conifold over a base space with SU(2) holonomy group. Taking the coordinates ($\theta,\phi,\psi$), we will
study a metric:
\begin{equation}
 ds^{2}_{10}=H^{-\frac{1}{4}}(r)\eta_{\mu\nu}dx^{\mu}dx^{\nu}+H^{\frac{3}{4}}(r)ds^{2}_{4},
\end{equation}
where
\begin{equation}
ds^{2}_{4} = \alpha^{2}(r)dr^{2}+r^{2}(\beta^{2}(r)\sigma^{2}+\gamma^{2}(r)(d\theta^{2}+sen^{2}\theta d\phi^{2}))
\end{equation}

\begin{equation}
 \sigma=d\psi + \cos \theta d\phi.
\end{equation}
The vielbeins of this metric are:

\begin{eqnarray}
 e^{1}=\alpha dr & , & e^{2}=r\beta\sigma \nonumber\\
 e^{3}=r\gamma d\theta & , & e^{4}=r\gamma sen \theta d\phi.
\end{eqnarray}
Following Carlevaro and Isr\"{a}el \cite{Carlevaro:2009jx}, we can take a set of complexified vielbeins of the form

\begin{eqnarray}
 E^{1}=e^{1}+ie^{2} & , & E^{2}=e^{3}+ie^{4},
\end{eqnarray}
and we can define the fundamental 2-form $J_{2}$ by

\begin{eqnarray}
 J_{2} & = & \frac{i}{2}\sum_{a=1}^{2}{A_{a}E^{a}\wedge \bar{E}^{a}}\nonumber\\
       & = & A_{1}e^{1}\wedge e^{2}+A_{2}e^{3}\wedge e^{4},
\end{eqnarray}
where $A_{1}$ and $A_{2}$ are constants. Therefore

\begin{equation}
 J_{2}\wedge J_{2} = A_{1}A_{2} e^{1}\wedge e^{2}\wedge e^{3}\wedge e^{4}.
\end{equation}
As $J_{2}\wedge J_{2}$ is proportional to the volume of $M_{4}$ then it satisfies the Strominger equation \cite{Strominger:1986uh}:

\begin{equation}
 d(e^{-2\phi}J_{2}\wedge J_{2})=0.
\end{equation}
Since
\begin{equation}
 dJ_{2}=\left(A_{1}r\alpha\beta + 2rA_{2}\gamma(\gamma + r \gamma')\right)sen\theta dr \wedge d\theta \wedge d\phi,
\end{equation}
the condition for $M^{4}$ be a K\"{a}hler manifold is

\begin{eqnarray}
dJ_{2}=0 & \Leftrightarrow & A_{1}r\alpha\beta + A_{2}2\gamma(\gamma'+r\gamma)=0.
\end{eqnarray}
We can follow \cite{Cvetic:2000mh} and take $A_{1}=-A_{2}=1$, so

\begin{equation}
 J_{2}= e^{1}\wedge e^{2}-e^{3}\wedge e^{4}.
\end{equation}
In this background, let us consider only the actions of the instanton term, what means $\star_{10}B_{2}=\star_{10}F_{2}=0.$
Therefore,
\begin{equation}
 d(\star_{10}\mathcal{H}_{3})=0.
\end{equation}
Now, let us search for solutions of the form \cite{Cvetic:2000mh}:

\begin{equation}
 e^{-2\phi}\star_{10}\mathcal{H}_{3}=dH^{-1}\wedge d^{6}x.
\end{equation}
Following the steps done in \cite{Cvetic:2000mh} we can relate the 2-form gauge $F_{2}$ with a self-dual 2-form $L_{2}$ as

\begin{eqnarray}
 F_{2} & = & mL_{2}\nonumber\\
       & = & mr^{n}(e^{1}\wedge e^{2}+e^{3}\wedge e^{4}).
\end{eqnarray}

In the ansatz above we have supposed a spherical symmetry and a power radial dependence only, whose exponent we will
determine
using the condition that $M^{4}$ is asymptotically Euclidean and that $L_{2}$ is a squared-integrable form \cite{Cvetic:2000mh,cglp}.
Then using the ansatz for $\mathcal{H}_{3}$ and $F_{2}$ the Bianchi identity yields the differential equation for the
warp factor

\begin{equation}
 \bigtriangleup_{4} H(r)= -m^{2}r^{2n}.
\end{equation}
This is a Poisson like equation for the warp factor where the Laplacian is evaluated over the $M^{4}$ and so

\begin{equation}
 \frac{d}{dr}\left(\frac{r^{3}\beta\gamma^{2}}{\alpha}H'(r)\right)-m^{2}r^{2n+3}\alpha\beta\gamma^{2}=0.
\end{equation}

Although the metric factors that determine the initial manifold
have singularities, we can find solutions of equation above that for a specific power $n$ it does not have singularity
anymore.

Let us now study the equation for the warp factor in a particular example of a Ricci-flat manifold.

\section{Resolution over Eguchi-Hanson spaces}
\begin{figure}
  \label{fatorwarp}
  \centering
 \includegraphics[width=0.7\textwidth]{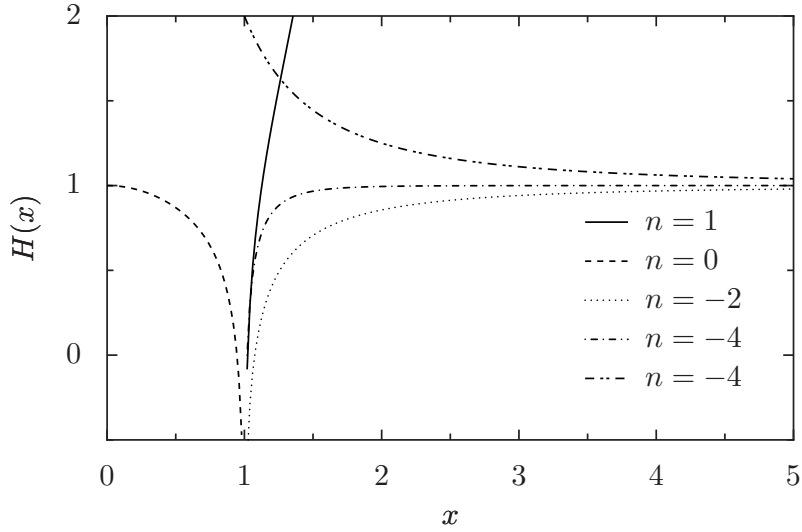}
  \caption{Graphic of warp factor for various powers $n$ for $a=1$. For $n>0$ the warp factor diverges
both at $r=1$ as at infinity.
For $a=0$ there is only interior
solution diverging at $a=1$ too. For negatives powers we have reached a well behaved solution only for $n\geq -4$ when we can find a integration constant that cancels
the logarithmic divergence.}
\end{figure}

As done in \cite{Cvetic:2000mh}, let us choose $M^{4}$ as the Eguchi-Hanson space. The factors of metric are:
\begin{eqnarray}
 \alpha^{2}=W(r)^{-1} & , & \beta^{2}=\frac{1}{4}W(r)\nonumber\\
 \gamma^{2}=\frac{1}{4} & , & W(r)=1-\frac{a^{4}}{r^{4}},
\end{eqnarray}
where $r\geq a$. This manifold is asymptotically locally Euclidean (ALE) with topology of the form
$S^{2}/\mathbb{Z}_{2}$ or
$\mathbb{C}^{2}/\mathbb{Z}_{2}$ and near $r=a$ the topology is $\mathbb{R}^{2}\times S^{2}$, with a bolt singularity.
Such factors satisfy the equation for closure of fundamental form $J_{2}$, so the Eguchi-Hanson space has a K\"{a}hler form given
above. The differential equation for warp factor is

\begin{equation}
 (r^{3}W(r)H'(r))'=-m^{2}r^{2n+3},
\end{equation}
whose solution is

\begin{equation}
 H(r)=C_{2}+\frac{m^{2}}{(2n+4)}\int{\frac{r'^{2n+5}}{(r'^{4}-a^{4})}}dr'+C_{1}\int{\frac{r'}{(r'^{4}-a^{4})}dr'}.
\end{equation}
As we are interested in solutions asymptotically Euclidean we can choose $C_{2}=1$ and must choose the power $n$ such
that $\lim_{r\rightarrow \infty}{H(r)}=1$. If $n\geq 0$, the first integral does not have an upper bound and so
diverges at
infinity. For $n=-1$, the manifold has a singularity at $r=a$ and diverge at infinity. For $n=-2$, we have the warp
factor
defined in the range $0\leq r \leq a$ only, having a singularity at $r=a$. For $n\leq-3$, $H(r)$ is defined only for
$a\leq r\leq \infty$, with $\lim_{r\rightarrow \infty}{H(r)}=1$. However, for $n=-3$ we still have a singularity in
$r=a$,
but for $n\leq -4$, i.e., $2n+5=-3$, as done in \cite{Cvetic:2000mh}, the power of r in the integrand is $p\leq -3$ and so the solution is:

\begin{equation}
 H(r)=1+\frac{m^{2}+a^{4}b}{4a^{6}}\ln\frac{(r^{2}-a^2{})}{(r^{2}+a^{2})}+\frac{m^{2}}{2a^{4}r^{2}}.
\end{equation}
There is a logarithmic divergence at $r=a$ but we can choose a
particular integration constant, $b=-\frac{m^{2}}{a^{4}}$ and
\begin{equation}
 H(r)=1+\frac{m^{2}}{2a^{4}r^{2}},
\end{equation}
so that the naked singularity in $r=a$ is taken off. In Fig. (\ref{fatorwarp}) we have plotted the
warp factor for various powers.


Using the ansatz for $\mathcal{H}_{3}$ field and the warp factor above we obtain:

\begin{equation}
 \mathcal{H}_{3}= -\frac{m^{2}}{2a^{4}}\frac{1}{r^{3}}W(r)^{\frac{1}{2}}e^{\theta}\wedge e^{\phi}\wedge e^{\sigma}.
\end{equation}
This solution is quite similar to that found by Carlevaro and Isr\"{a}el \cite{Carlevaro:2009jx} in six dimensions. From this field we
can define the 2-form of torsion:

\begin{equation}
 T^{a}=\frac{1}{2}H^{a}_{bc}e^{b}\wedge e^{c}.
\end{equation}

It worthwhile to point out that the tensor $\mathcal{H}^{a}_{bc}$ can be interpreted as a non-metric connection \cite{kms,Becker:2003yv}. Since $\mathcal{H}^{a}_{bc}$ is skew-symmetric we have a torsion
even though the geometric connection for the Eguchi-Hanson spaces is symmetric.

Hence, the components of torsion are

\begin{eqnarray}
T^{1}      & = 0\\
T^{2} & = \frac{m^{2}}{a^{4}}r^{-5}H^{-\frac{3}{4}} W^{\frac{1}{2}} e^{3}\wedge e^{4}\\
T^{3} & = \frac{m^{2}}{a^{4}}r^{-5}H^{-\frac{3}{4}} e^{2}\wedge e^{4}\\
T^{4} & = \frac{m^{2}}{a^{4}}r^{-5}(\sin{\theta})^{-2}H^{-\frac{3}{4}} e^{2}\wedge e^{3}
\end{eqnarray}

\begin{figure}
  \label{torsion1}
  \centering
 \includegraphics[width=0.7\textwidth]{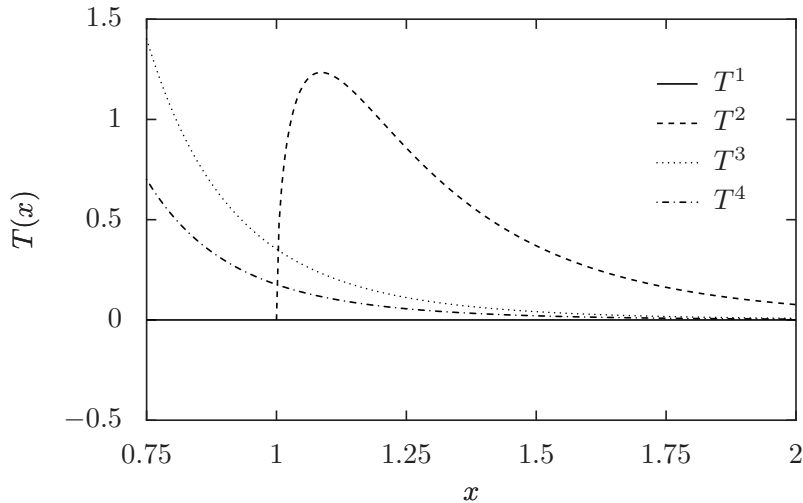}
  \caption{Graphic of components of torsion 2-form for $a=1$. The component $T^{1}$ is identically null while the $T^{3}$ and $T^{4}$ components diverge at origin and
vanish in UV regime. The $T^{2}$ component is defined only for $r\geq 1$ vanishing in this point and at infinity. Note that there is a neighborhood around $r=1$ where
the torsion is non-null.}
\end{figure}

In Fig. (\ref{torsion1}) we have plotted the graphic for the torsion above for $a=1$. The last two above are defined for whole $r>0$, diverging in $r\rightarrow 0$ and going quickly to zero when $r\rightarrow \infty$.
The $\sigma$ component is defined only for $r>a$ diverging in this point and vanishing at infinity. So the torsion diverges in the IR limit ($r\rightarrow 0$) and vanish in the UV limit ($r\rightarrow \infty$).

\section{Resolution through BF term}

Now we are interested in the resolution through a $BF$ term. We will use an ansatz
for $B$ field previously used in Ref. \cite{Klebanov:2000hb}, that have spherical symmetry and are defined only at spherical cycle:

\begin{equation}
 B_{2}=f(r)e^{\theta}\wedge e^{\phi}
\end{equation}
This ansatz was studied by Klebanov and Tseytlin \cite{Klebanov:2000nc} in the context of conformal invariance of gauge-gravity duality on conifolds.
A necessary condition is \cite{Klebanov:2007us}:

\begin{equation}
 4\pi^{2}\left(\frac{1}{g_{1}^{2}}-\frac{1}{g_{2}^{2}}\right)=\frac{1}{g_{s}e^{\phi}}\left(\frac{1}{2\pi\alpha'}\int_{S^{2}}{B_{2}-\pi}\right).
\end{equation}
The equations for $\mathcal{H}_{3}$ are:

\begin{eqnarray}
 d(\star \mathcal{H}_{3}) & = & -k\star F_{2}\\
 d\mathcal{H}_{3} & = & B_{2}\wedge F_{2}.
\end{eqnarray}

Let us to extend the ansatz chosen in \cite{Cvetic:2000mh} to

\begin{equation}
 e^{-\phi}\star_{10}\mathcal{H}_{3}= (dH^{-1}+ \omega_{1})\wedge dx^{6},
\end{equation}
where the 1-form $\omega_{1}$ is such that
\begin{equation}
 d\omega_{1}\wedge dx^{6}=-\star_{10}F_{2}.
\end{equation}
This is a system of differential equations relating the components of $\omega_{1}$ to $F_{2}$ and so it has a solution.
This ansatz yields the equation
\begin{equation}
 d\star_{4}dH + d\star_{4} \omega_{1}= B_{2}\wedge F_{2}.
\end{equation}

Note that in this limit we have changed only the transgression of Bianchi identity compared with \cite{Cvetic:2000mh}.
Assuming the same ansatz previously used for $F_{2}$ we have

\begin{equation}
 B_{2}\wedge F_{2}= -mf(r)r^{n}e^{r}\wedge e^{\sigma}\wedge e^{\theta}\wedge e^{\phi}.
\end{equation}
For $||B_{2}||>>||F_{2}|| \Rightarrow d\omega_{1}\rightarrow 0 $ and this yields

\begin{equation}
 \bigtriangleup H(r)=-mf(r)r^{n}.
\end{equation}
Let us take a power-dependence for f, like $f(r)=r^{k}$, thus,
\begin{equation}
 \frac{d}{dr}\left(\frac{r^{3}\beta\gamma^{2}H'(r)}{\alpha}\right)=-m\alpha\beta\gamma^{2}r^{n+k+3}.
\end{equation}
Taking the Eguchi-Hanson space as a background for this defect,
we obtain the differential equation for $H$ as

\begin{equation}
 \left(r^{3}W(r)H'(r)\right)'=-mr^{n+k+3},
\end{equation}
whose solution is:

\begin{equation}
 H(r)=1+\frac{m}{(n+k+4)}\int{\left[\left(\frac{r'^{n+k+5}}{r'^{4}-a^{4}}\right)+c_{1}\frac{r'}{(r'^{4}-a^{4})}\right]dr'}.
\end{equation}
%
As seen before, there is a resolved solution for $n+k+5\leq -3$. Since $||B|| >>||F||$ we can take the powers
$n=-10$ and $k=2$, for instance. Therefore

\begin{equation}
 H(r)=1+\frac{m^{2}}{2a^{4}r^{2}}.
\end{equation}
Although the resolution is given by the BF term instead of instanton one we have reached the same warp factor. Since
the NS-NS field strength is given by the warp factor the torsion has the same expression for the resolution obtained using the instanton
defect that diverges only at origin. However, if we evaluate the non-topological NS-NS field strength using the ansatz for $B_{2}$ we obtain

\begin{equation}
 B_{2}=r^{2}e^{\theta}\wedge e^{\phi} \Rightarrow H_{3}=dB_{2}=2rW^{\frac{1}{2}}(r)e^{r}\wedge e^{\theta}\wedge e^{\phi}.
\end{equation}
This ansatz yields a torsion in the form

\begin{eqnarray}
T^{1} & = rW(r)^{\frac{3}{2}}H(r)^{-\frac{3}{4}}e^{3}\wedge e^{4}\\
 T^{2} & = 0\\
 T^{3} & =r^{-1}H^{-\frac{3}{4}}(r)e^{1}\wedge e^{4}\\
 T^{4} & =r^{-1}H^{-\frac{3}{4}}(r)e^{1}\wedge e^{3}.
\end{eqnarray}
The component in r direction is defined only for $r>a$, diverging at infinity. The other directions are defined for all r vanishing
in origin and diverging in UV regime. Since the whole topological NS-NS field strength $\mathcal{H}_{3}$ vanishes in UV limit we conclude that the
topological 3-form added to $H_{3}$ cancels the diverging components resulting in an asymptotically torsion free manifold.

\section{Conclusions}
We have studied the process of resolution of a conifold by an instanton and a BF term. Since both
belongs to same Chern character that appears in the anomaly term in type II string theories we heuristically have proposed the same
change $\mathcal{F}_{2}=B_{2}+kF_{2}$ in the heterotic theory. Since $k$ is related with the string scale $l_{s}$, the BF term is
relevant for distances of string scale and the instanton term is relevant for distances greater than the string scale.
The presence of the BF term yields a current proportional to $\star_{10}F_{2}$ and alter the Bianchi identity including
the same $B_{2}\wedge F_{2}$ besides of $F_{2}\wedge F_{2}$. For the instanton term there is no current associated
with this term and so we can choose an ansatz relating to dual of $\mathcal{H}_{3}$ with the derivative of warp factor.
This ansatz yields a Poisson differential equation relating the laplacian to the instanton charge. Taking the Eguchi-
Hanson space as background, we have found solutions previously studied by Cvetic, L\"{u} and Pope \cite{Cvetic:2000mh} that do not
have singularities and have a non-vanishing torsion that diverges in IR regime ($r\rightarrow 0$) and vanishes in UV regime ($r\rightarrow \infty$),
so asymptotically the manifold is torsion free. For the BF term we have a current for $\mathcal{H}_{3}$ proportional
to the dual of $F_{2}$ and so we must to add a new term in the differential equation for the warp factor. Assuming
$||B_{2}||\gg ||F_{2}||$ we can disregard this term and solve the Poisson equation previously studied. Since the warp factor
is the same as in the instanton case the BF term resolves the singularity and has the same torsional behavior as instanton
term. Evaluating only the $dB_{2}$ component of $\mathcal{H}_{3}$ we have found
a solution with asymptotically divergent torsion and we conclude that the remainder topological term cancels the divergence.

Nevertheless, we have reached a solution for the $||B_{2}||\gg||F_{2}||$ case where the term
$d\omega_{1}$ is negligible and we could ask if a resolved solution can be reached for $||B_{2}||\approx ||F_{2}||$ and
$||B_{2}||\ll||F_{2}||$. Moreover, we have studied only cases $B_{2}\wedge F_{2}$ and $F_{2}\wedge F_{2}$ separately and we could study a limit where both
defects are working. We have limited us to an Abelian gauge group and it could be interesting to extend this work to full
$E_{8}\times E_{8}$ or $SU(32)$ groups as done in \cite{Carlevaro:2008qf}. Another feature is to study the resolution of BF term
taking into
account the Gauss-Bonnet term $tr(R_{2}\wedge R_{2})$ in transgression of Bianchi identity. We could also
extend this resolution taking 6-dimensional spaces instead of 4-dimensional ones, using the Strominger equations
to find the (1,1)-fundamental form $J_{1,1}$ and the (3,0)-holomorphic form $\Omega_{3,0}$ as done in \cite{Carlevaro:2009jx} for
the instanton term. To do it, we must to extend the DUY (Donaldson-Uhlenbeck-Yau) equations, conditions for stability of the gauge bundle, to the BF
term. In another perspective one can also consider the present study in M-theory following the lines of Ref. \cite{Brito:2002xv}, where was pointed out that
Chern-Simons corrections associated with anomaly on the M5-brane can resolve singularity on M2-brane.

Going to AdS-CFT correspondence, since the BF term resolves the conifold, an interesting question is to study the
renormalization group flow under the BF term, for instance. In the context of AdS-QCD duality, since BF term resolves the
singularity as well as the instanton term, we can ask what kind of fields are holographically related to BF term.

The authors would like to thank Funda\c{c}\~{a}o Cearense de apoio ao Desenvolvimento
Cient\'{\i}fico e Tecnol\'{o}gico (FUNCAP) and Conselho Nacional de Desenvolvimento
Cient\'{\i}fico e Tecnol\'{o}gico (CNPq) and CAPES-PROCAD for financial support.

\end{document}